\documentclass[twocolumn,nofootinbib,
showpacs,pra,aps,superscriptaddress,floatfix]{revtex4}

\usepackage{graphicx}

\begin{document}
\def\la{\langle}
\def\ra{\rangle}
\def\ome{\omega}
\def\ome0{\omega_0}
\def\om0{\omega_0}
\def\Om{\Omega}
\def\vep{\varepsilon}
\def\wh{\widehat}
\def\P0{\wh{\cal P}_0}
\def\dt{\delta t}

\newcommand{\beq}{\begin{equation}}
\newcommand{\eeq}{\end{equation}}
\newcommand{\beqa}{\begin{eqnarray}}
\newcommand{\eeqa}{\end{eqnarray}}
\newcommand{\intf}{\int_{-\infty}^\infty}
\newcommand{\into}{\int_0^\infty}

\title{Tunneling dynamics in relativistic and nonrelativistic
wave equations}

\author {F. Delgado}
\email{qfbdeacf@lg.ehu.es}
\affiliation{Departamento de Qu\'{\i}mica-F\'{\i}sica,
UPV-EHU,\\
Apartado 644, 48080 Bilbao, Spain}
\author {J. G. Muga}
\email{qfpmufrj@lg.ehu.es}
\affiliation{Departamento de Qu\'{\i}mica-F\'{\i}sica,
UPV-EHU,\\
Apartado 644, 48080 Bilbao, Spain}
\author{A. Ruschhaupt}
\email{rushha@physik.uni-bielefeld.de}
\affiliation{Departamento de Qu\'{\i}mica-F\'{\i}sica,
UPV-EHU,\\
Apartado 644, 48080 Bilbao, Spain}
\author{G. Garc\'{\i}a-Calder\'on}
\email{gaston@fisica.unam.mx}
\affiliation{Instituto de F\'{\i}sica,
Universidad Nacional Aut\'onoma de M\'exico\\
Apartado Postal {20 364}, 01000 M\'exico, D.F., M\'exico}
\author{J. Villavicencio}
\email{villavics@uabc.mx}
\affiliation{Facultad de Ciencias,
Universidad Aut\'onoma de Baja California\\
Apartado Postal 1880, 22800 Ensenada, Baja California, M\'exico}

\begin{abstract}
We obtain the solution of a relativistic wave equation and compare it 
with the solution of the Schr\"odinger equation for
a source with a sharp onset and excitation frequencies below 
cut-off. 
A scaling of position and time reduces to a single 
case all the (below cut-off) nonrelativistic solutions,  
but no such simplification holds for the relativistic equation,
so that 
qualitatively different ``shallow'' and ``deep''
tunneling regimes may be identified relativistically.
The nonrelativistic forerunner
at a position beyond the penetration length of the  
asymptotic stationary wave does not tunnel; nevertheless, it   
arrives at the traversal 
(semiclassical or B\"uttiker-Landauer) time $\tau$.  
The corresponding relativistic
forerunner is more complex: it oscillates due to the interference between 
two saddle point contributions, and may be  
characterized by two times for the arrival of the maxima of lower and upper
envelops.  
There is in addition 
an earlier relativistic forerunner, right after the causal front, 
which does tunnel.  
Within the penetration length, tunneling is more robust 
for the precursors of the relativistic equation. 
\end{abstract}


\maketitle

\section{INTRODUCTION}

Tunneling is a general feature of wave equations: 
for frequencies below the ``cutoff'',  wavenumbers become imaginary 
and the waves evanescent. Even so, it has  
been mainly discussed 
for the Schr\"odinger equation due to the   
shocking contrast between  
the behavior of classical and quantum particles.  
It has been also traditionally examined for  
stationary conditions, until a paper by    
B\"uttiker and Landauer \cite{BL82} triggered the interest  
on its temporal aspects,
again, mostly within the framework of the Schr\"odinger
equation. 
  
Tunneling dynamics may be very counterintuitive when compared with 
the evolution of propagating waves. 
Defining ``tunneling times'' has been in particular  
non trivial in quantum theory (Ref. \cite{TQM} is a recent and extense multi-author
review; for previous reviews see \cite{HS89,LA90,LM94,Chiao98,Ghose99}).
After twenty years of discussions and many proposals 
since \cite{BL82}, 
it is clear that different time scales are
relevant depending on the experiment, wave feature, or 
observable quantity investigated. There is thus not a single
tunneling time scale, but several; many of them may    
be obtained, related to each other, and classified systematically
by means of a theory  
that exploits the non-commutability of the 
operators implied \cite{BSM94}.   

One of the most important and commonly found scales 
is the ``traversal'', ``semiclassical'' or 
``B\"uttiker-Landauer'' time $\tau$. 
For the dimensionless Schr\"odinger equation for a particle moving in a 
constant potential, 
\beq
i\frac{\partial \psi}{\partial t}=-\frac{1}{2}
\frac{\partial^2 \psi}{\partial x^2}+\psi,
\label{Sch}
\eeq
(all quantities are dimensionless throughout this work
except in the discussion of section \ref{III}),
and for a frequency $\omega_0<1$, it is given by \cite{BL82} 
\beq\label{taunr}
\tau=\frac{x}{\kappa_0},
\eeq
where 
\beq
\kappa_0=[2(1-\om0)]^{1/2} 
\eeq
in Eq. (\ref{taunr}) plays the role of a semiclassical ``velocity'' which 
increases for deeper and deeper tunneling  
(i.e. for smaller $\om0$). 
$\tau$ is the characteristic time for the transition from 
sudden to adiabatic regimes in oscillating barriers, and for the 
spin rotation in a weak magnetic field. A recent and surprising 
finding 
is that it is also the time of 
arrival of the peak of the forerunner  
that appears, beyond the penetration length
of the asymptotic stationary wave,
$x>1/\kappa_0$, 
for a ``source'' with a sharp onset, 
\begin{equation}
\psi(x=0,t)= e^{-i\omega_0 t}\Theta(t),
\label{source}
\end{equation}
and
\begin{equation}
\psi(x>0, t<0)=0.    
\label{source1}
\end{equation}
$\Theta(t)$ is the Heaviside function; in the language of electromagnetic 
wave propagation it is the envelop function of
the input pulse, 
whereas  $\omega_0$ is the ``signal'', ``carrier'',
or ``excitation'' frequency. 
More precisely, the arrival time is $t_p^t=\tau/3^{1/2}$ if the peak 
of the density $|\psi(x,t)|^2$ is 
evaluated with respect to $t$ for $x$ fixed \cite{MB00}, and 
$t_p^x=\tau$ if it is evaluated with respect
to $x$ for $t$ fixed \cite{VRS02}. 

This role of $\tau$ is indeed surprising because that
forerunner {\it is not tunneling} beyond $1/\kappa_0$, where it is 
predominantly composed by over-the-barrier ($\omega>1$) components  
\cite{MB00}.
The importance of $\tau$ is blurred however within the penetration shell, 
$x<1/\kappa_0$, where  the arrival of the tunneling peak 
obeys a different time scale, inversely proportional to $\kappa_0^2$ rather 
than to $\kappa_0$ \cite{gcv01,GVDM02}.
In the small-$x$ region the arrival time of the peak  
$t_p^t$ as a function of $x$ is 
rather flat in comparison with the 
(linear in $x$) arrival time of the peak in the large-$x$ region. 
It actually forms a basin with a minimum, 
so that the precursor may arrive later at smaller values of $x$
\cite{gcv01,GVDM02}.   
Moreover the forerunner {\it does tunnel} at small $x$. This is another 
unexpected  result \cite{GVDM02}, after 
so many works where precursors had been always related to 
an above-the-barrier passage \cite{RMA91,TKF87,APJ89,BM96}.
Surely no attention had been paid to the small-$x$ region because the  analytical 
approximations based on saddle and pole contributions of the integral defining the 
solution fail, as discussed below. This region is the easiest to observe though.

Our initial motivation to undertake the present work was to determine if
and how  
the above recent results extend to the 
relativistic case. Especifically, our aim is to investigate  
the role played by $\tau$, and the characteristics of the 
forerunners 
at large  
and small $x$ for under-cut-off, unit-step modulated excitations 
with the relativistic wave equation  
\beq
\left[\frac{\partial^2}{\partial t^2}-\frac{\partial^2}{\partial x^2}
+1\right]\psi=0,   
\label{KGdl}
\eeq
which is much more accessible experimentally (with 
waveguides in evanescent conditions \cite{Nimtz,RRSAS01,RM02})
than the Schr\"odinger equation. These are all aspects that previous 
studies of relativistic Klein-Gordon equations with evanescent 
conditions had not 
explored \cite{DL93,JLL01}. Some preliminary, partial results 
on the role of the relativistic $\tau$
may be found in \cite{BT,MB00}.      

Note that the stationary
equations corresponding to Eqs. (\ref{Sch}) and (\ref{KGdl}) are
equal and have equal solutions, but the two  
cases are not equivalent in the time domain, because the
dispersion relations
between $\omega$ and $k$ are different, 
\beqa
\omega^2&=&1+k^2\;\;\;\;\;\; ({\rm relativistic}), 
\label{di1}\\
\omega&=&1+k^2/2\;\;  ({\rm non\,\, relativistic}). 
\label{di2}
\eeqa
%
In fact we have found a wealth of 
qualitative changes with respect to the Schr\"odinger scenario
as the following sections will show.  
One of them is discussed in section \ref{II}, namely, the absence for 
the relativistic equation of the
scaling properties that simplify the nonrelativistic solutions; 
this implies that  qualitatively
different solutions are possible
for 
different carrier frequencies in the relativistic case.    
Section \ref{III} deals with possible physical realizations of the 
formalism.  
The technical tools used are an analytical series expression for the exact 
solution, which is obtained in section \ref{IV}, and 
the asymptotic analysis of section \ref{V}, based on 
previous work by B\"uttiker and Thomas \cite{BT},
that will allow to describe  
and understand the wave behaviour at ``large-$x$'' in simple terms. 
For small-$x$ this analysis is substituted by numerical 
exploration in section \ref{VI},
except in the proximity of the very first front 
or ``causal limit'' at $t=x$, where a simple analytical approximation
is again available.    

\section{Scaling properties\label{II}}

The solution of Eqs. (\ref{KGdl}) or (\ref{Sch}) 
with the boundary condition of Eq. (\ref{source})
is given in both cases by  
\beq\label{solu}
\psi(x,t;\omega_0)=\frac{i}{2\pi}
\intf d\omega\, \frac{e^{ikx-i\omega t}}{\omega-\om0+i0},
\eeq
as may  be checked by substitution.
We shall only consider $x>0$. 
It is easy to see from Eq. (\ref{solu}) that
the wave vanishes for $t<x$.   
In the relativistic case  $k=(\omega^2-1)^{1/2}$
is defined with two branch cuts from 
$\omega=\pm 1$ downwards in the complex $\omega$-plane,
as shown in Figure 1, 
whereas for the nonrelativistic equation, 
$k=[2(\omega-1)]^{1/2}$ is defined with only one branch cut
downwards from $\omega=1$. 
For evanescent waves, i.e., when $k=i\kappa$, 
$\kappa>0$ the relativistic wavenumber tends to the nonrelativistic one
as $\omega\to 1$,    
\beq
\kappa=(1-\omega^2)^{1/2}\to_{\omega\to 1}[2(1-\omega)]^{1/2}. 
\eeq

In the nonrelativistic case it is useful to shift the frequency axis
by defining the variable 
\beq
\Omega=\omega-1  
\eeq
(in particular $\Omega_0=\om0-1$). Let us also define  
$\phi(x,t;\Omega_0)\equiv e^{it}\psi(x,t;\omega_0)$.  
If we introduce a scale factor 
\beq
\Omega=\alpha\Omega',\;\; \alpha>0,
\eeq
it follows from Eq. (\ref{solu}) that 
\beq\label{scaling}
\phi(x,t;\Omega_0)=
\phi(\alpha^{1/2}x, \alpha t;\Omega_0/\alpha),      
\eeq
which means, in words, that any two solutions with excitation frequency  
above cut-off ($\Omega_0>0$) are related to each other 
by the scaling law of Eq. (\ref{scaling}) and similarly,  
all solutions below cut-off ($\Omega_0<0$) are also
simply related to each other by scaling.
We can generate all possible solutions from 
two of them, one below and one above cut-off.   
Nevertheless, Eq. (\ref{scaling}) does not hold 
for the relativistic wave equation because of the different 
dispersion relation,
so that solutions for different excitation frequencies $\omega_0$
cannot reduce to each other; 
they may and do change qualitatively when sweeping over $\ome0$, as
demonstrated in Section \ref{IV}.

\section{Physical realizations\label{III}}

Even though we use throughout the paper dimensionless
variables and equations, it is worth noting that Eq. (\ref{KGdl})
leads to  
the Klein-Gordon equation for spin-0 particles by means of 
the substitutions
\beqa
x&=&X(m_0c)/\hbar,
\\
t&=&Tm_0c^2/\hbar,
\eeqa
where $X$ and $T$ denote dimensional position and time, 
and $m_0$ and $c$ are the particle's rest mass and the velocity of light
in vacuum. 
Another interesting connection, much more promising to  
implement the present analysis experimentally \cite{RM02},  
and free from the conceptual puzzles
of the former,  may be established  
with the equation that governs the electromagnetic field
components in waveguides
of general constant cross section with perfectly
conducting walls \cite{Kristensson95}. 
This requires the substitutions  
\beqa
x&=&X\lambda,
\\
t&=&T\lambda c,
\eeqa
where $\lambda>0$ is one of the eigenvalues 
of the waveguide.  

Dimensional Schr\"odinger equations 
corresponding to Eq. (\ref{Sch}) may be obtained 
formally in the limit $c\to\infty$ (the proper 
one from the Klein-Gordon equation for particles,
and an analogous one 
from the waveguide equation). Physically, the solution 
of the dimensionless wave equation
approaches the solution of the dimensionless 
Schr\"odinger equation when it is dominated
by $\omega$-components close to one (the cut-off), note that
the relativistic dispersion relation, Eq. (\ref{di1}), tends to the
nonrelativistic one, Eq. (\ref{di2}), as $\omega\to 1$.      
Indeed  
excitations with $\ome0\approx 1$ 
provide relativistic solutions very close
to the nonrelativistic ones  
except  
in the region   
near 
the causal limit $t=x$ (the ``first precursor''),
where the whole frequency range 
of the source signal has a significant influence. 

In order to compare relativistic and nonrelativistic equations 
the same ``excitation'' will be used, 
Eq. (\ref{source}), which for the waveguide may be understood as a 
compact representation of (real) sine and cosine excitations.
Since for the relativistic equation the excitations
$e^{\pm i\ome0 t}\Theta(t)$ produce at $x$ the complex 
conjugate wave of each other, only the interval 
$0<\ome0<1$ will be examined.  
Also, to carry out the comparison with the Schr\"odinger equation
in terms of analogous quantities,    
we shall study $|\psi|^2$ which, in the relativistic-waveguide 
case may be simply regarded 
as the sum of the squares of the waves resulting from
the sine and cosine 
excitations. Because of its meaning in the Schr\"odinger 
equation, we shall refer to this quantity as the ``density'', even 
in the relativistic case.   
The agreement between the relativistic and nonrelativistic 
solutions near the cut-off and far from $t=x$,  
combined with the scaling property of Eq. (\ref{scaling}) may 
serve in practice to simulate in a waveguide
the solution of the Schr\"odinger equation for any $\omega_0$ in the time 
domain. 
\begin{figure}
{\includegraphics[angle=-90,width=2.8in]{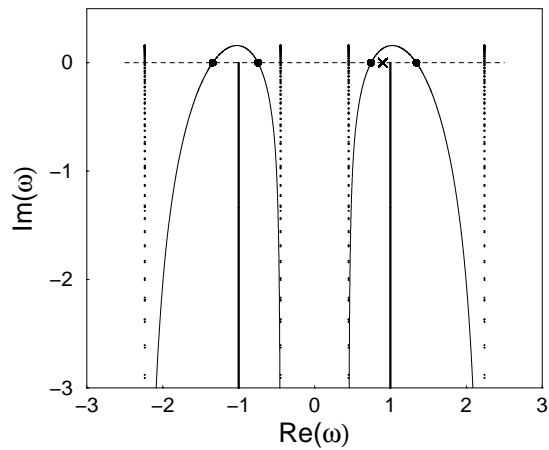}}
\caption[]{Steppest descent paths in the complex $\omega$ plane for $t=1.5$,
$\ome0=0.85$ and $x=1$ (narrow solid line). The thicker solid 
lines are the two branch cuts while the
verticals lines are the asymptotes. The dots mark $\pm\omega_s$
and $\pm1/\omega_s$; the cross is at $\omega_0$.}
\label{figure0}
\end{figure}

\section{Relativistic series solution\label{IV}}

In this section we shall obtain the solution of Eq. (\ref{KGdl})
in series form. It will be used to provide exact numerical 
results as well as analytical approximations   
%
%
for $x,t>0$ and for the sharp onset source of 
Eq. (\ref{source}),  
%
%
following refs.  
\cite{gcv99,vjpa00}. 
We begin by Laplace transforming the equation (\ref{KGdl})
using the standard definition,
\begin{equation}
\overline{\psi}(x;s)=\int\nolimits_{0}^{\infty }\psi (x,t)e^{-st}dt.
\label{v3}
\end{equation}
The Laplace transformed solution reads, 
\begin{equation}
\overline{\psi}(x;s)=c_1e^{-px},\qquad x\geq 0,  
\label{v4}
\end{equation}
where we have defined $p=(s^2+1)^{1/2}$. The corresponding
Laplace transform 
of Eq. (\ref{source}), yields,
\begin{equation}
\overline{\psi}(0;s)=\frac{1}{s+i\omega_0}.  
\label{v5}
\end{equation}
By combining Eqs. (\ref{v4}) and (\ref{v5}) evaluated at $x=0$ we can
determine the value of the constant $c_1$, 
\begin{equation}
\overline{\psi}(x;s)=\frac{e^{-px}}{s+i\omega_0},\qquad x\geq 0.
\label{v6}
\end{equation}
The time dependent solution for $x>0$ is obtained by performing the
inverse Laplace transform of Eq. (\ref{v6}), using the Bromwich
integral formula,
\begin{equation}
\psi (x,t)=\frac{1}{2\pi i}\int\limits_{\gamma ^{\prime }-i\infty }^{\gamma
^{\prime }+i\infty }\overline{\psi}(x;s)e^{st}ds, \qquad t>0,    
\label{v7}
\end{equation}
where the integration path is taken along a straight line $s=\gamma ^{\prime
}$ parallel to the imaginary axis in the complex $s$-plane. The real
parameter $\gamma ^{\prime }$ can be chosen arbitrarily as long as all
singularities remain to the left-hand side of $s=\gamma ^{\prime }$.

To avoid dealing with the branch points at $s=\pm i$ in Eq. (\ref{v7}), let us
introduce the change of variable, $-iu=(s+p)$. Thus, $p=i(u^{-1}-u)/2$, and the 
integral may be written as
%
%
\begin{equation}
\psi (x,t)=\frac 1{2\pi i}\int\limits_{i\gamma -\infty }^{i\gamma +\infty
}\Phi(u)du,  
\label{v8}
\end{equation}
where the new integrand $\Phi(u)$ is given by
\begin{eqnarray}
\Phi(u) &=&\frac{(1-u^2)}{u(u-\omega_0-k_0)(u-\omega_0+k_0)}  \nonumber \\
&&\times \text{exp}\{i[u(x-t)-u^{-1}(x+t)]/2\}, 
\label{v9}
\end{eqnarray}
and $k_0=k(\omega_0)$. 
The integrand has now an essential singularity at $u=0$ and
two simple poles $u_{\pm}=(\omega_0 \pm k_0)$.  
For $x>t$ we close the integration path 
from above, by a large semicircle $\Gamma _1$ of radius $R$, forming a
closed contour $C_1$. The contribution along $\Gamma _1$ vanishes as $%
R\rightarrow \infty $, and since there are no poles enclosed inside $C_1,$ $%
\psi (x,t)=0$ for $x>t$. For the case $x<t,$ we close the integration path
from below with a large semicircle $\Gamma _2$. The closed contour $C_2$
contains three small circles $C_{0,}$ $C_{+}$ and $C_{-}$ enclosing the
essential singularity at $u=0$ and the simple poles at $u_{+}$ and $u_{-},$
respectively. Hence, it follows that,
\begin{equation}
\frac 1{2\pi i}\left[ \int\limits_{i\gamma -\infty }^{i\gamma +\infty
}-\int\limits_{\Gamma
_2}+\int\limits_{C_0}+\int\limits_{C_{+}}+\int\limits_{C_{-}}\right]
\Phi(u)du=0.
\label{v10}
\end{equation}
The integrals corresponding to the contours $C_+$  and $C_-$ can be easily
evaluated, 
\begin{equation}
-\frac {1}{2\pi i}\int\limits_{C_{\pm }}\Phi(u)du = 
e^{(\pm k_0x-i\omega_0 t)}.
\label{v11}
\end{equation}
The contour integration for $C_0$ involves an essential singularity at $u=0$.
Introducing 
the change of variable
$v =-iu\xi^{-1}$ the
integral becomes
\begin{equation}
-\frac {1}{2\pi i}\int
\limits_{C_0}\Phi(u)du=\frac {1}{2\pi i}\int
\limits_{C_0^{^{\prime }}}\frac{1}{\xi^2}
\frac{(1+v^2 \xi^2) e^{\eta (v -v^{-1})/2}}{v(v -v_+)(v -v_{-})}dv,
\label{v12}
\end{equation}
where
\beqa
v_{\pm}&=&z_{\pm}/i\xi,
\\
z_{\pm}&=&(\omega_0 \pm k_0),
\\ 
\xi &=&\left[
(t+x)/(t-x)\right]^{1/2},
\\
\eta &=&(t^2-x^2)^{1/2}.
\eeqa
The 
integration is carried out by first
separating the integrand into partial fractions,
and then substituting
the formula for the Bessel generating function,
\begin{equation}
e^{\eta (v -v^{-1})/2}=\sum_{n=0}^\infty v^n J_n(\eta
)+\sum_{n=1}^\infty (-1)^n v^{-n} J_n(\eta ),
\label{v13}
\end{equation}
and the series expansion,
\begin{equation}
(v_{\pm}-v)^{-1}=(v_{\pm })^{-1}\sum_{n=0}^\infty (v
/v_{\pm })^n.  
\label{v14}
\end{equation}
Now we can calculate the resulting integrals by means of the residue theorem.
For the case of an essential singularity, the residue may be determined by
computing explicitly the coefficient corresponding to $v^{-1}$ from
the series expansion and their products. In that case, equation (\ref{v12})
becomes,
\begin{eqnarray}
-\frac 1{2\pi i}\int\limits_{C_0}\Phi(u)du &=&\left[ 
-J_0(\eta )-\sum_{n=1}^\infty (-1)^n\frac{J_n(\eta)}{(v_+)^n}%
\right.  \nonumber \\
&&\left. -\sum_{n=1}^\infty (-1)^n\frac{J_n(\eta)}{(v_-)^n}%
\right].
\label{v15}
\end{eqnarray}
Finally, by feeding the results given by Eqs. (\ref{v11})
and (\ref{v15}) into 
Eq. (\ref{v10}), the solution for the internal region reads
\begin{equation}
\psi(x,t)=\left\{ 
\begin{array}{ll}
\psi(x,k_0,t)+\psi(x,-k_0,t), & t>x \\ 
0, & t<x,
\end{array}
\right.
\label{stepfinal}
\end{equation}
with $\psi(x,\pm k,t)$ defined as
\begin{eqnarray}
\psi(x,\pm k_0,t)&=& e^{(\pm ik_0x-i\omega_0 t)}+\frac{1}{2}%
J_0(\eta )\nonumber \\
&&-\sum\limits_{n=0}^\infty (\xi /iz_{\pm })^nJ_n(\eta ) .
\label{simplif}
\end{eqnarray}
%
It is worthwhile to remark the similarity of the above solution with that
of Eq. (16) of ref. \cite{gcv99} and Eqs. (15) and (16) of ref. \cite{vjpa00}.
{}From Eq. (\ref{v13}),
\begin{equation}
\sum\limits_{n=0}^\infty (\xi /iz_{\pm })^n J_n(\eta)=
e^{i[\pm k_0x-\omega_0 t]}
-\sum\limits_{n=1}^\infty (-1)^n (iz_\pm /\xi)^n J_n(\eta).
\end{equation}
Therefore, Eq. (\ref{simplif}) may also be written in the form 
\begin{equation}
\psi(x,\pm k_0,t)= \frac{1}{2}J_0(\eta )
+\sum\limits_{n=1}^\infty (-1)^n (iz_\pm/\xi)^n J_n(\eta ), 
\label{pre}
\end{equation}
which will be useful for the analysis of the
region close to the causal  
limit $t=x$. 

The above relativistic solutions should be contrasted with the solution to 
the nonrelativistic equation using the same initial condition \cite{MB00,VRS02},
\begin{equation}
\psi(x,t)=\frac{1}{2}{\rm e}^{-it}{\rm e}^{ix^2/(2t)}[w(y_{k_0})+w(y_{-k_0})]
\label{nrel}
\end{equation}
where $w(y_{\pm k_0})$ stands for the ``$w$-function'' \cite{as} with  argument $y_{\pm k_0}$,
\begin{equation}
y_{\pm k_0}=i{\rm e}^{-i\pi/4}\left( \frac{1}{2t}\right)^{1/2}\left[x\mp k_0 t\right].
\label{arg}
\end{equation}

\section{Saddle-pole approximation for large $x$\label{V}}

%
%

Following Sommerfeld and Brillouin \cite{SB},
who studied the propagation of a unit step-function modulated signal 
in a Lorentz medium\footnote{For a modern 
and more accurate treatment see \cite{OS94}},  
the forerunners
and other wave features
may be understood and quantify with asymptotic analysis 
techniques.  
Let us 
deform the integral of Eq. (\ref{solu}) along the 
steepest descent paths (SDP), see Figure \ref{figure0}, defined by 
\beq
\omega_I^\pm=\frac{-(\pm\omega_R\omega_s-1)(\pm\omega_R-\omega_s)}
{[(\omega_s^2-1)
(-\omega_R^2\pm2\omega_R\omega_s-1)]^{1/2}}
\eeq
($\omega_R$ and $\omega_I$
are the real and imaginary parts of 
$\omega$),
where the upper sign is for the positive saddle at $\omega_s$ and the lower sign
for the negative saddle\footnote{Compare
with the only nonrelativistic saddle
at $\omega_s=1+x^2/(2t^2)$, which is obtained from the 
positive relativistic saddle for $t>>x$.} at $-\omega_s$, 
\beq
\label{oms}
\pm \omega_s=\pm\frac{t}{(t^2-x^2)^{1/2}},  
\eeq
plus a clockwise circle around the pole at $\om0$ that must be added to the
contour if the pole has been crossed by the SDP. The SDP 
have asymptotes at $\omega_R=\pm[\omega_s\pm (\omega_s^2-1)^{1/2}]$ 
and cross the real axis at $\pm \omega_s$ and $\pm \omega_s^{-1}$,  
see figure \ref{figure0}. 
The crossing of the pole occurs when $\omega_s^{-1}=\om0$, 
i.e., for fixed $x$, at time 
\beq
\tau=\frac{x}{\kappa_0}=\frac{x}{(1-\om0^2)^{1/2}},
\eeq
which generalizes the traversal time of Eq. (\ref{taunr}) 
for the relativistic case \cite{BT}.     
Note that $\tau>x$, i.e., the crossing occurs always after the 
arrival of the first front imposed by  
relativistic causality at $t=x$.  
One might be tempted to believe that the crossing of the pole 
is associated with the arrival of a ``monochromatic'' $\om0$-front
\cite{Stevens}.  
That this is not the case has been pointed out in several works on the
Schr\"odinger equation, the reason
being the dominance of over-the-barrier components
associated with the saddle. 
The frequency analysis of the relativistic
forerunner is more complex as the following
discussion will demonstrate.

Applying the standard asymptotic approximation for 
the saddle contributions \cite{BH86}, 
$\psi$ takes the form
\beqa
\psi(x,t)&\approx& \psi_p + \psi_s^+ + \psi_s^-,
\label{apro}\\
\psi_p&=&\Theta(t-\tau)e^{-\kappa_0 x}e^{-i\om0 t}, 
\\
\psi_s^\pm&=&i\sqrt{\frac{\mp i x^2}{2\pi t^2}}
\frac{\omega_s}{\omega_s\mp\omega_0}\exp[\mp i\eta].
\label{sadd}
\eeqa
%
%
%
Eq. (\ref{apro}) is a good approximation as
long as the ``width'' of the 
saddle is small compared to the ``distance'' from the saddle 
to the pole \cite{BT},  
\beq
(\omega_s\mp \omega_0)^2
\frac{\partial^2\phi}{\partial \omega^2}\bigg|_{\omega=\omega_s}
=\frac{\eta^3}{x^2}(\omega_s\mp \omega_0)^2>>1,
\eeq
where $\phi=\omega t-kx$. This always happens for large enough $t$ 
and fails very close to $t=x$.   
Applying this condition at $t=\tau$ we get
\beqa
x\om0\kappa_0&>>&1\;\; ({\rm positive\; saddle}), 
\label{c+}
\\
x\om0(1+\om0^2)^2/\kappa_0^3&>>&1\;\; ({\rm negative\; saddle}),
\label{c-}
\eeqa
to be compared with the simpler nonrelativistic criterion
\cite{MB00} 
\beq
x\kappa_0>>1.
\eeq

At variance with the nonrelativistic case, the condition 
of Eq. (\ref{c-}) may only worsen with decreasing $\om0$, 
whereas 
the one in Eq. (\ref{c+})
improves from $\om0=1$ up to the critical value   
$\om0=(1/2)^{1/2}\approx 0.7$, but also worsens below. 
$\psi_s^+$ is dominant with respect to $\psi_s^-$
as $\om0\to 1$ and for large t
but the importance of the negative saddle increases with decreasing 
$\om0$ and t, 
\beq
\frac{|\psi_s^+|}{|\psi_s^-|}=\frac{t+\eta \ome0}{t-\eta \ome0}.
\eeq
In particular, at $t=\tau$, 
\beq
\frac{|\psi_s^+|}{|\psi_s^-|}=\frac{1+\om0^2}{1-\om0^2}.
\eeq
To compare the relative importance of the saddle and pole contribution
we can examine the ratio $\left|\psi_p/\psi_s^+\right|$, 
%
\beqa
{\cal R}(x,t)&=& \frac{|\psi_p|}{|\psi_s^+|}=
\pi e^{-2 \kappa_0 x}\frac{t^2\eta}{x^2}
\nonumber\\
&\times&\left(1-
\frac{\ome0\eta}{t^2} \right) \Theta(t-\tau).
\label{posa}
\eeqa
${\cal R}(t,x)$ grows monotonously with $t$ and decreases with $x$. 
The time scale for the attainment of the stationary regime, or equivalently,
the duration of the transient regime dominated by the saddles before the
pole dominates, $t_{tr}$, can be identified as the time when ${\cal R}=1$.
Assuming that $\tau \ll t_{tr}$, this time is given by 
\beq
\label{ttr}
t_{tr}=\frac{e^{{2\kappa_0 x}/{3}}x^{2/3}}{(2\pi)^{1/3}(1-\ome0)^{2/3}}, 
\eeq
which is an exponentially large quantity as in the
Schr\"odinger equation
\cite{MB00}. Indeed, Eq. (\ref{ttr}) tends to the nonrelativistic result 
taking 
$\ome0 \to 1$.

\begin{figure}
\rotatebox{-90}
{\includegraphics[height=8cm]{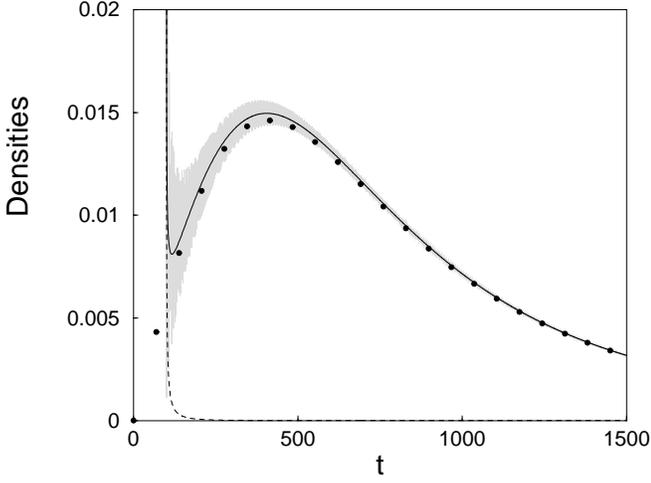}}
\caption[]{
$|\psi_s^+|^2$(solid black line), $|\psi_s^-|^2$ (dashed line) and
$|\psi_s^++\psi_s^-|^2$ (solid grey line); 
$|\psi_s|^2$ (nonrelativistic: dots). $x=100$ and $\ome0=0.99$.}
\label{fig4}
\end{figure}

\subsection{``Densities''}
If the pole contribution is negligible compared to the two saddle
point contributions, the total density is given by
\beqa
&&|\psi(x,t)|^2\approx|\psi_s^++\psi_s^-|^2=
\cr\cr &&\frac{t^2+\ome0^2\eta^2+x^2(t^2/\tau^2+\ome0^2)\sin(2\eta)}
{\pi\eta x^2\left(t^2/\tau^2+\ome0^2\right)^2},\;\;{\cal R}\ll 1
\label{modul}
\eeqa

\begin{figure}
\rotatebox{-90}{\includegraphics[height=8cm]{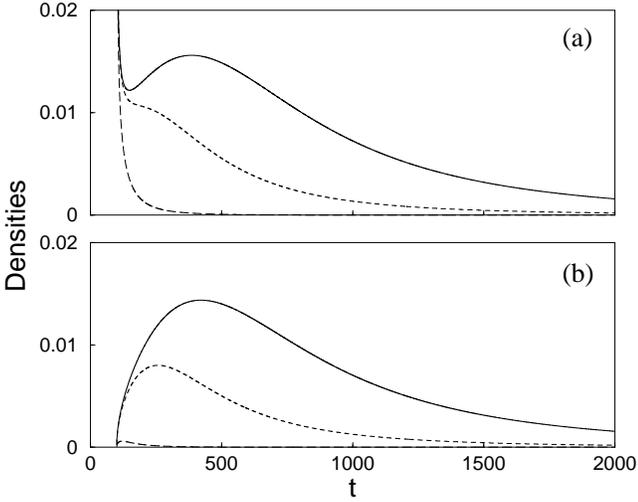}}
\caption[]{Upper (a) and lower (b) envelops at  
at $x=100$ for : $\ome0=0.99$ (solid
line), $\ome0=0.97$ (dashed line) and $\ome0=0.5$ (long-dashed line).}
\label{fig1}
\end{figure}

\begin{figure}
\rotatebox{-90}{\includegraphics[height=8cm]{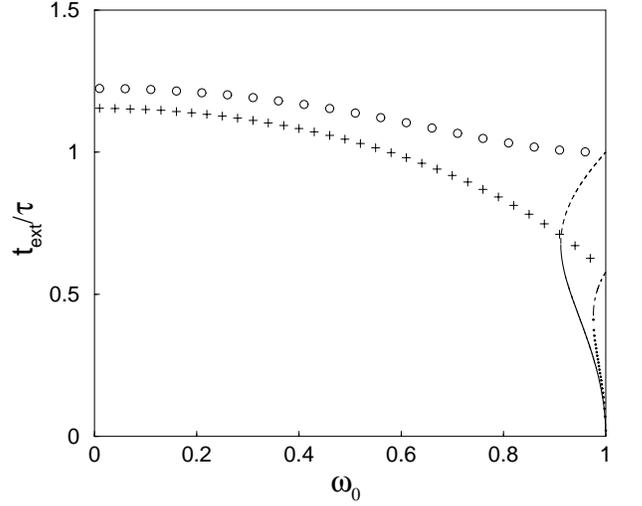}}
\caption[]{
$t_{max}^{x,U}/\tau$ (dashed line);  
$t_{min}^{x,U}/\tau$ (solid line);  
$t_{max}^{t,U}/\tau$ (dotted-dashed line);  
$t_{min}^{t,U}/\tau$ (dotted line); 
$t_{max}^{x,L}/\tau$ (circles), and 
$t_{max}^{t,L}/\tau$ (crosses)}. 
\label{rsop}
\end{figure}

The interference between the two saddles 
leads to an oscillatory pattern of the density, whose amplitude
$A$ increases for 
deeper tunneling, see Figure \ref{fig4}. 
{}From Eq. (\ref{modul}), 
\beq
A= \frac{x^2}{\pi\eta(t^2-\ome0^2\eta^2)}.
\label{ampli}
\eeq
The oscillation only disappears in the nonrelativistic 
limit $\omega_0\to 1$, $t>>x$. 

Since the dominant frequencies of the saddle terms for 
$\psi_s^+$ and $\psi_s^-$ are $\pm\omega_s$ respectively, 
the oscillation period of
the interference pattern between the two saddles is very well
approximated by $\pi/\omega_s$.  
%
%
Of course,  beyond $t_{tr}$ the oscillations
fade away and are substituted by the asymptotic dominance of the pole term.  

The upper and lower envelops of $|\psi_s^++\psi_s^-|^2$ 
are given, from Eq. (\ref{modul}), by:
\beqa
&&|\psi_s^++\psi_s^-|^2_{U}=\frac{2t^2}{\pi x^2\eta
\left(t^2/\tau^2+\ome0^2 \right)^2},
\cr
&&|\psi_s^++\psi_s^-|^2_{L}=\frac{2\ome0^2\eta}{\pi x^2\left(t^2/\tau^2
+\ome0^2\right)^2}.
\label{envol}
\eeqa
%
%
A second forerunner, after the first one at $t=x$, may thus 
be identified thanks to the maxima of these envelops, see Fig. \ref{fig4}. 
The upper 
envelop may hold a maximum and a minimum, but they disapear below
a critical value of the injection frequency $\ome0$, see Fig. \ref{fig1} a. 
We may thus clearly distinguish between shallow and 
deep tunneling regimes with qualitatively different features for the 
upper envelop. 
On the other hand,
the lower
envelop has only a maximum that remains for any $\ome0$, although its 
amplitude decreases with smaller    
$\ome0$, see Eq. (\ref{envol}) and Fig. \ref{fig1} b, 
and becomes physically irrelevant in comparison to the upper envelop. 

Taking the derivative with respect to $t$ or $x$ of the envelops,
we can extract 
``temporal'' or ``espacial extrema'' respectively. In 
the case of the lower envelop, 
\beqa
t_{max}^{x,L}&=&\frac{\tau}{2}\left[{3-\ome0^2+\sqrt{9(1+\ome0^4)
-14\ome0^2}}\right]^{1/2},
\nonumber\\
t_{max}^{t,L}&=&\left(\frac{4-3\ome0^2}{3}\right)^{1/2}\tau,
\eeqa
which are always larger than the corresponding 
nonrelativistic limits, $\tau$ and $\tau/\sqrt{3}$ respectively. 
For the upper envelop we obtain
\beqa
t_{max/min}^{t,U}&=&\frac{\tau}{\sqrt{6}}\left[2-\ome0^2\pm
\sqrt{4-28\ome0^2+
25\ome0^4}\right]^{1/2},
\nonumber\\
t_{max/min}^{x,U}&=&\frac{\tau}{2}\left[1+\ome0^2\pm
\sqrt{1-22\ome0^2+25\ome0^4}\right]^{1/2}.
\eeqa
%

The ratios between the above critical times and $\tau$, 
$t_{ext}/\tau$ are represented in Figure \ref{rsop}. None of these 
ratios depend on $x$. 
This is explained  
by an interesting scaling property of the 
envelops (which is also satisfied separately by $|\psi_s^\pm|^2$),   
%
\beqa
&\alpha&|\psi_s^+(\alpha x,\alpha t)+\psi_s^-(\alpha x,\alpha t)|^2_{U/L}
\nonumber
\\&=&
|\psi_s^+(x,t)+\psi_s^-(x,t)|_{U/L}^2
\eeqa
which holds for any fixed excitation energy and 
has been also noted in the nonrelativistic 
saddle contribution to the density \cite{VRS02}.
A consequence is that the maxima and minima  
of the envelops  
with respect to $x$ and $t$ 
travel at constant velocity, which may only depend on $\om0$.
Nevertheless, when the interference is taken into account 
a similar relation is not satisfied by Eq. (\ref{modul}). 
Equation (\ref{oms}) implies that $\omega_s$, and 
therefore the oscillation period of the density oscillations, remain 
constant when $x$ and $t$ are scaled, instead of increasing as 
the  
time 
span of the forerunners does.      
%

\begin{figure}
{\includegraphics[angle=0,width=3.in]{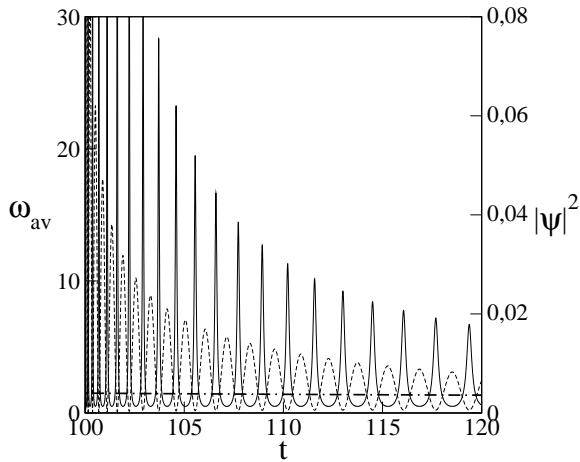}}
\caption[]{$|\psi|^2$ and $\omega_{av}$ versus time for 
$x=100$ and $\ome0=0.5$ (dashed line and solid line respectivity).
Also shown is the nonrelativistic $\omega_{av}$ (dotted-dashed line).}
\label{master}
\end{figure}

\subsection{Instantaneous frequency}
Figure \ref{master} shows a typical
probability density for
the first forerunner region at large $x$,  
where the two saddle contributions dominate clearly over
the pole contribution.
We have choosen a signal frequency $\ome0=0.5$ and a fixed position 
$x=100$ ($\tau \approx 710$).
In this scale, there are no differences between the exact 
and the aproximate solution in Eq. (\ref{apro}) except
in the very limit $t=x$.
In the same figure 
we have plotted the average local instantaneous 
frequency $\omega_{av}\equiv
-Im[\left(\partial{\psi}/\partial t\right)/\psi]$
versus time \cite{Cohen95,MB00}.
Since now $\psi\approx\psi^+_s+\psi¯_s$ we obtain,  
\beq
\omega_{av}=\frac{\ome0\left(-2t^2y+x^2\cos(2y)\right)}{\eta
\left[-t^2-\ome0^2\eta^2+
(-t^2+\ome0^2\eta^2)\sin(2y)\right]}, 
\label{wav}
\eeq
with envelops 
\beq
\omega_{av}^{U}=\frac{t^2}{\ome0\eta^2},\;\;\;\;
\omega_{av}^{L}=\ome0. 
\eeq
The density and $\omega_{av}$ are dephased. In particular 
the maxima of the density correspond to the 
minima at $\omega_0$ of the frequency,  so that 
most of the density corresponds to frequencies below the cut-off.
(This effect is more and more clear  
for lower injection frequencies.)
This means that the first forerunner 
is essentially tunneling whereas, in the same time domain, the nonrelativistic 
wave is not. For the second forerunner, however, the minima of the density 
are not so close to zero, see e.g. Figure \ref{fig4},
so that there is an alternating influence
of frequencies above and below cut-off.  

%
%

\section{Exact solution for small $x$\label{VI}}

\begin{figure}[h]
{\includegraphics[angle=0,width=3.5in]{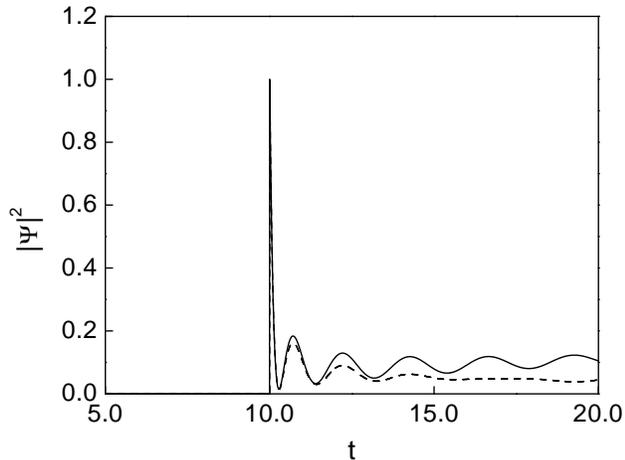}}
\caption[]{$|\psi|^2$ versus time at $x=10$ and $\ome0= 0.99$ for the exact calculation 
(solid line) and the first term approximation (dashed line).}
\label{wavefr}
\end{figure}

The small $x$ region may be defined by the 
failure of the inequalities in Eqs. (\ref{c+}) and (\ref{c-}). 
It requires in general an exact numerical treatment
because the simple saddle-pole aproximation is not 
valid anymore. Figure \ref{wavefr} exhibits a plot of the ``density'' as a function of time,
for given values of $x$ and $\omega_0$, near  $\eta  \approx 0$
using Eq.\ (\ref{pre}). The exact calculation (solid line)  exhibits a sharp relativistic wavefront, reaching unity at $x=t$, followed by mild oscillations of smaller amplitude and eventually by the asymptotic stationary density. Figure \ref{wavefr} provides also a comparison with the one-term approximation (dashed line) to Eq.\ (\ref{pre}), namely,
\begin{equation}
\psi(x,t) \approx J_0(\eta) -2i \frac{\omega_0}{\eta}J_1(\eta).
\label{aprox}
\end{equation}
One sees that the above approximation gives an excellent description in the vicinity of the relativistic wavefront. 

\begin{figure}[h]
\rotatebox{-90}{\includegraphics[height=8cm]{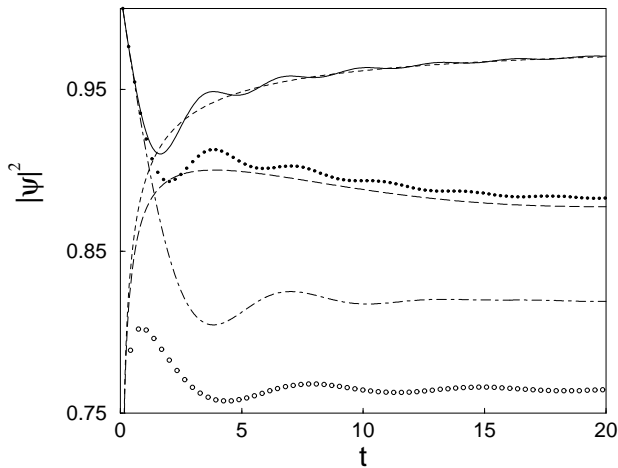}}
\caption[]{$|\psi|^2$ versus time at $x=0.1$ for three differents
values of $\ome0$:
$\ome0= 0.99$ (solid line), $\ome0=0.8$ (dotted line) and $\ome0=0.1$
(dotted-dashed line). The same quantity 
is plotted  for the nonrelativistic solution (dashed line, $\ome0=0.99$;
long dashed line, $\ome0=0.8$,  and circles for $\ome0=0.1$).}
\label{fig5}
\end{figure}

In Figure \ref{fig5}, we have plotted the ``density''
versus time for three different values of $\ome0$. The first forerunner 
just after $t=x$ may be  characterized from the  expansion of the  two Bessel 
functions in Eq. (\ref{aprox}) for very small $\eta$ \cite{as},   
\beq
\psi(x,t) \approx 1-(t-x)\left[-i\frac{\ome0}{2}-\frac{t+x}{4}\right].
\label{limit}
\eeq

\begin{figure}[h]
\rotatebox{-90}{\includegraphics[height=8cm]{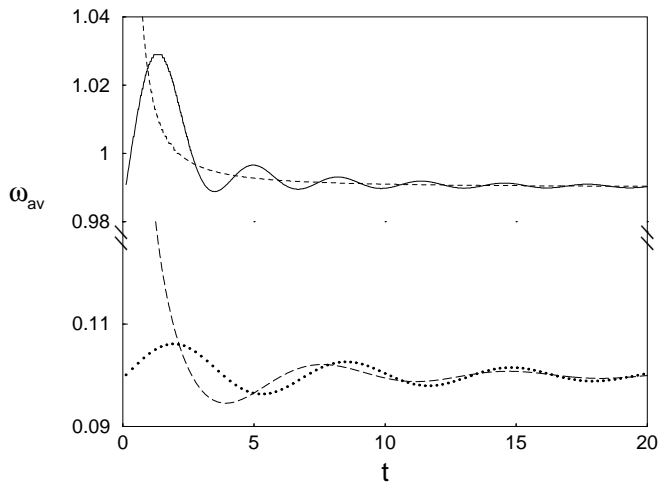}}
\caption[]{$\omega_{av}$ versus time at $x=0.1$ for two differents
values of the injection frequency:
$\ome0= 0.99$ (solid line), $\ome0=0.1$ (dotted line). The same quantity 
is plotted  for the nonrelativistic solution (dashed line)
for $\ome0=0.99$ and
long dashed line for $\ome0=0.1$. The axis has been cut to show all curves  
with the same scale.}
\label{fig6}
\end{figure}

\begin{figure}[h]
\rotatebox{-90}{\includegraphics[height=8cm]{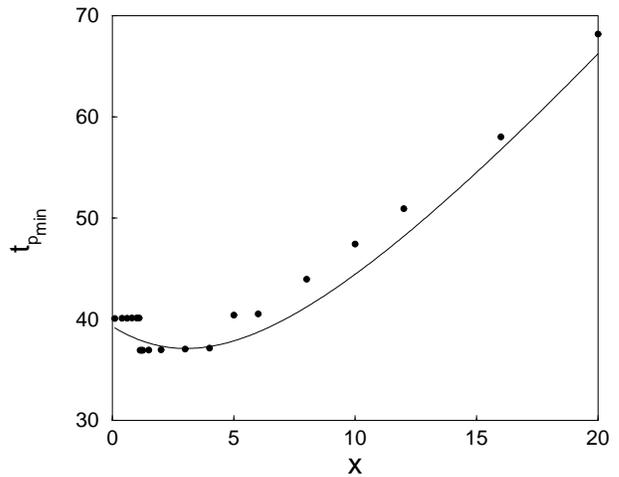}}
\caption[]{time of arrival of the maximum of the lower envelop
for the relativistic case (dots) and for the first forerunner in the 
nonrelativistic case (solid line). $\ome0$ has been choosen equal to 0.98.
}
\label{fig9}
\end{figure}

\begin{figure}[h]
\rotatebox{-90}{\includegraphics[height=8cm]{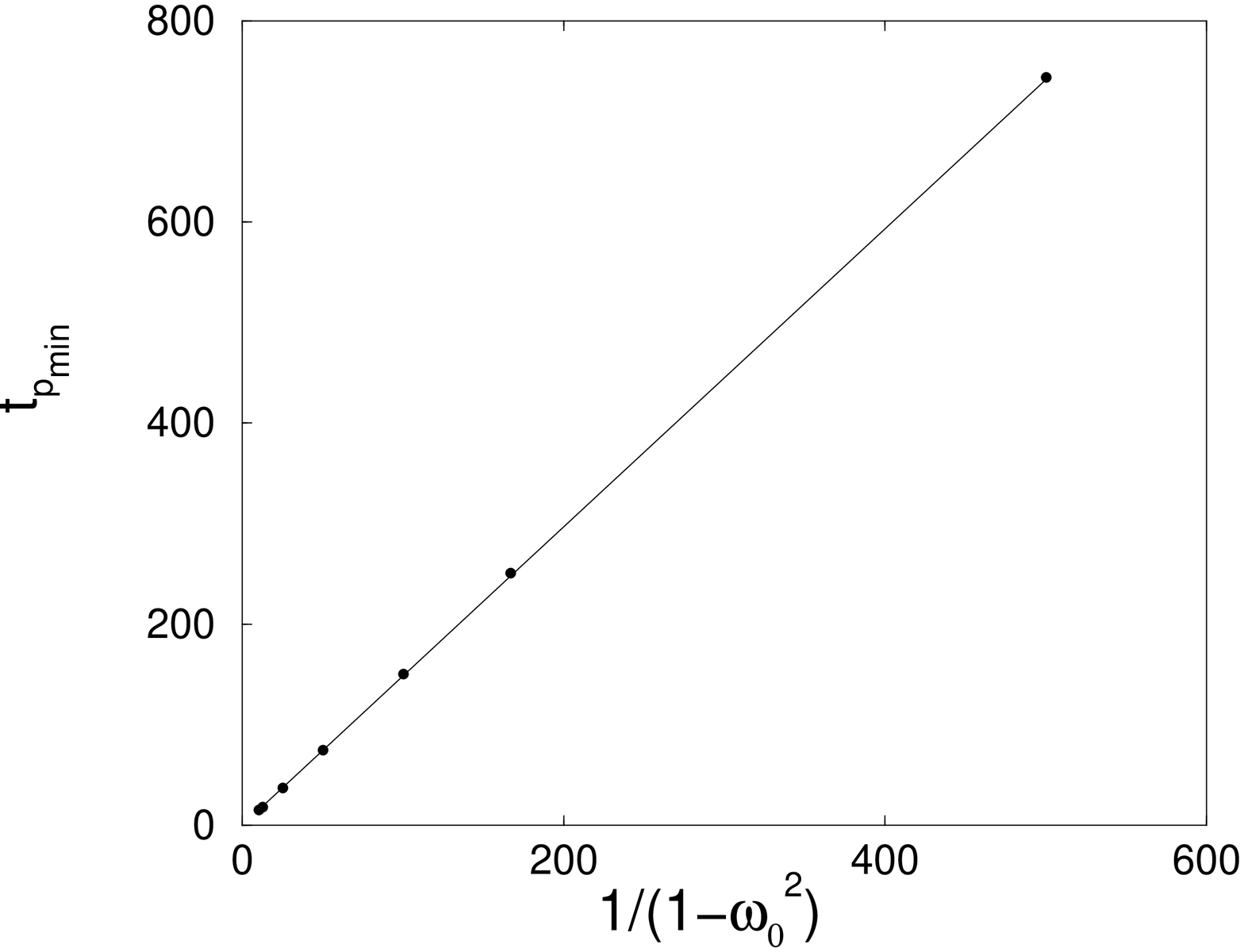}}
\caption[]{$t_{p_{min}}$ versus $1/\kappa_0^2$ using $\ome0$ from 0.95 to 
0.999. }
\label{fig7}
\end{figure}

It consists of a peak that decays from 1 with slope $-(t+x)/2$
independently of $\ome0$. At variance with the nonrelativistic equation, 
that  immediately smoothes away any initial singularity, the relativistic
equation propagates the sharp jump  
of the  excitation in the input signal.    
Note in Figure \ref{fig5} that for
deeper tunneling the asymptotic 
relativistic signal becomes stronger than the non
relativistic one:  
the pole contribution to the density has the same {\it form}
in both cases \cite{MB00},
$e^{-2\kappa_0}$, but the $\kappa_0$'s are only equal 
at $\omega_0=0$. Otherwise,  
$\kappa_0$ ({nonrelativistic}) $>\kappa_0$ ({relativistic})
for $0<\omega_0<1$. 

In fact, the approximation in Eq. (\ref{limit}) is valid for
any value of $x$, but its relevance is mostly appreciated at 
small $x$. At large $x$ the first peak is immediately 
followed by rapid oscilations of large amplitude, 
so that the first forerunner is characterized better by the upper envelop
in that case, except in the immediate proximity of the causal limit.   

The average local instantaneous frequency at the causal
limit $x=t$ is always $\ome0$, 
see Figure \ref{fig6} and Eq. (\ref{limit}), which is quite 
different from the ``high'' frequencies of the  
nonrelativistic case, see Figure \ref{fig6} 
($\omega_{av} (t=x;$ {nonrelativistic}$)=1.5$ for $\omega_0=0.99$).     
%
In fact the first precursor as well as the rest of the wave  
tunnel cleanly at any time if the source frequency is small enough. 
This limit value of
$\ome0$ for total tunneling decreases when $x$ increases.  

We may define as before 
the second forerunner according to some convention. 
Numerically it is relatively simple to identify 
the maximum of a lower envelop for  
$\ome0\gtrsim0.8$.   
Figure \ref{fig9} shows that, as in the nonrelativistic case,
we have a region where
the signal arrives at earlier times at greater $x$,  
but for large $x$ 
it grows linearly with $x$. 
The arrival time of
the envelop's peak at the minimum of the basin,  
$t_{p_{min}}$,
depends linearly on  $1/\kappa_0^2$, see Figure \ref{fig7},
as in the nonrelativistic
case.   

\section{Summary and discussion}

In this work we have described the forerunners and 
the transition to the asymptotic regime  
of the solution of a relativistic wave equation 
for a unit-step-function modulated input signal 
with carrier frequencies below cut-off. 
This extends the investigation carried out previously for the
Schr\"odinger equation \cite{MB00,VRS02,GVDM02},
with the bonus that the relativistic equation 
may be physically implemented in 
waveguides. The main differences
between the relativistic and nonrelativistic cases are: (a) 
the relativistic solutions are not 
simply related to each other by time and position scaling as in the 
nonrelativistic case, so that qualitatively different 
``shallow'' and ``deep'' 
tunneling regimes may be distinguished;  (b) tunneling is more robust
relativistically, both in the precursors 
and asymptotically;
(c) the ``first'' relativistic precursor, right after the 
limit imposed by causality does not have a nonrelativistic counterpart, 
and does tunnel; (d) the ``second'' precursor, 
which tends to the nonrelativistic one for excitation
frequencies near cut-off, has an oscillating 
structure that may be characterized by its envelops. The traversal time 
$\tau$ is not an exact measure of their arrival except in the
nonrelativistic limit.

While the emphasis here has been on the comparison between 
the results of the Schr\"odinger and the relativistic
wave equation for a canonical input signal, our next objective 
is to incorporate to the theory more elements relevant for the 
waveguide experiments, such as dissipation, other forms of  
input envelop pulses (e.g. square or Gaussian),
waveguides with section constraints or barriers 
of lower dielectric constant \cite{Ermig96} 
(for  
classically forbidden regions of finite width,
the nonrelativistic description\cite{gcv01,gcv03} exhibits also the 
``shallow'' and ``deep'' tunneling regimes mentioned above),
frequency band-limitations, and a separate analysis of
cosine or sine excitations, which have been 
combined here to relate directly the relativistic 
wave amplitudes to  the nonrelativistic densities.   

\begin{acknowledgments}

FD and JGM and AR are grateful to I. L. Egusquiza for many discussions. 
They also acknowledge support 
by Ministerio de Ciencia y Tecnolog\'\i a (BFM2000-0816-C03-03), 
UPV-EHU (00039.310-13507/2001), and the Basque Government (PI-1999-28).
G-C and JV acknowledge financial support from DGAPA-UNAM under grant IN101301.
\end{acknowledgments}


\end{document}